# Dynamic Information Security Management Capability: Strategising for Organisational Performance

*Research in Progress*

**Mazino Onibere**
School of Computing and Information Systems
The University of Melbourne
Victoria, Australia
Email: mazino.onibere@unimelb.edu.au

**Atif Ahmad**
School of Computing and Information Systems
The University of Melbourne
Victoria, Australia
Email: atif@unimelb.edu.au

**Sean B Maynard**
School of Computing and Information Systems
The University of Melbourne
Victoria, Australia
Email: seanbm@unimelb.edu.au

## Abstract

The increasing frequency, impact, consequence and sophistication of cybersecurity attacks is becoming a strategic concern for boards and executive management of organisations. Consequently, in addition to focusing on productivity and performance, organisations are prioritizing Information Security Management (ISM). However, research has revealed little or no conceptualisation of a dynamic ISM capability and its link to organisational performance. In this research, we set out to 1) define and describe an organisational level dynamic ISM capability, 2) to develop a strategic model that links resources with this dynamic capability, and then 3) empirically demonstrate how dynamic ISM capability contributes to firm performance. By drawing on Resource-Based Theory (RBT) and Dynamic Capabilities View (DCV), we have developed the Dynamic ISM Capability model to address the identified gap. As we develop this research, we will empirically test this model to demonstrate causality between ISM capability and organisational performance.

# 1   Introduction

Information security has become a strategic concern for boards and senior management of organisations because of the increasing frequency, impact, consequence and sophistication of cyber security incidents (Ahmad et al. 2012; Ahmad et al. 2015; Baskerville et al. 2014). Now, not only must organisations contend with a business landscape that is characterised by competition, dynamic in nature and economical challenges; they must also contend with an information security landscape that is equally dynamic, rapidly evolving, characterised by highly complex and sophisticated threats. The cyber threats of today have become unpredictable, novel with a shift in emphasis on financial gain (Ahmad et al. 2019; Smiraus and Jasek 2011; Sood and Enbody 2013; Tankard 2011). Attackers are no longer the stereotypical hackers; rather, they are well funded organisational entities and nation states, who seek to steal information as part of economic or industrial espionage, and/or covertly sabotage critical infrastructure as part of cyber warfare (Ahmad et al. 2014; Desouza et al. 2020; Schiavone et al. 2014).

However, research has revealed an underdeveloped strategic perspective of the Information Security Management (ISM) function (Maynard et al. 2018). We found little or no conceptualisation of dynamic ISM capability and its link to organisational performance. In this study, we address the following research question:

> *How can Information Security Management enhance organisational performance?*

In this research, we look to the Strategic Management literature, where concepts of strategy, strategic management and firm performance are well established. We draw from both the resource-based theory (RBT) which argues that an organisation's resources form the basis of achieving competitive advantage; and the dynamic capabilities view (DCV) that introduces the concept of an unstable environment, thus, requiring that organisations have capabilities for dynamically reconfiguring their resource base. To address the research question, we seek to 1) define and describe an organisational level dynamic ISM capability, 2) to develop a strategic model that links resources with this dynamic capability, and then 3) empirically demonstrate how it contributes to firm performance

This research will make important contributions to both theory and practice. First, the proposed variance model strives to empirically demonstrate causality between ISM capability and firm performance, which is not currently discussed in ISM literature. Second, it extends existing ISM domain knowledge, which is currently very sparse in information security strategy, by drawing concepts from RBT and DCV and interpreting them in the context of ISM. This study contributes to practice. First, the proposed framework can help organisations recognise and develop their ISM capability, which eventually leads to better firm performance. Second, empirically demonstrated causality between ISM capability and firm performance can be used by organisations to strengthen their justification for investment in their ISM capability.

This paper is organised as follows. First, we discuss the theoretical concepts such as RBT, DCV and their relevance to firm performance from strategic management literature. Second, we interpret these concepts into the ISM discipline and develop the dynamic ISM capability model. Third, we define the perceived organisational value derived from ISM as a construct of firm performance and relate this to the dynamic ISM capability. Finally, we conclude the paper with future directions and expected contributions of the study.

# 2   THEORETICAL BACKGROUND

## 2.1   Resource Based Theory

The Resource Base Theory (RBT) argues that when firms possess valuable, rare, inimitable and non-substitutable (VRIN) resources, they can achieve competitive advantage (Barney 1991; Vanpoucke et al. 2014). However, as soon as these resources are copied or substituted, organisations lose their competitive advantage. According to RBT, resources are defined as assets, whether physical, geographical, human or organisational owned and/or controlled by the organisation and are inputs into organisational processes (Barney 1991; Dehning and Stratopoulos 2003; Eisenhardt and Martin 2000; Vanpoucke et al. 2014). Resources can be both internal and external, where external refers to such resources owned by vendors and partners that the organisation can access and utilise. Teece (2000) argues that the basis of competitive advantage is gradually shifting to knowledge assets, rather than physical assets, thus posing a different set of strategic and managerial challenges than before.

There is a lack of consensus on the definition of the various terminologies used in RBT such as resources, competences, core competences, capability, etc. While some authors have used resource as all-encompassing term, others have attempted to differentiate these terms. This research however subscribes to the approach adopted by Peppard and Ward (2004), which acknowledges a difference between resources, competences and capability, and recognises that organisational capability operates at the strategic and enterprise level.

RBT is rooted in a conception of a stable environmental context (Huang et al. 2015). RBT looks inwards into the organisation. It assumes that the external environment is relatively stable, and that the firm's performance is dependent on how the internal resources, competences, capabilities, etc. are utilised.

## 2.2 Dynamic Capabilities View

The DCV complements RBT by introducing the concept of an unstable and dynamic environmental context. DCV argues that in a dynamic and unstable environment, it is not enough to have and develop VRIN resources to sustain competitive advantage, firms must be able to reconfigure their existing base of resources in line with the changing environmental context. DC represent a firm's ability/capacity to reconfigure existing resources, in the face of significant industry and/or business change, to generate new value creating strategies – where reconfiguration may include acquisition, release, modification, integration, or recombination of resources (Eisenhardt and Martin 2000; Grant 1996; Pisano 1994). These DC may be codified in organisational processes or routines that are in place to change or augment existing resources in the face of anticipated change, or they may be purely experiential and specific to the unanticipated environmental change.

These strategic routines are complicated, detailed, analytic processes that rely extensively on existing knowledge and linear execution to produce predictable outcomes (Eisenhardt and Martin 2000). While strategic routines are useful in a moderately dynamic environment in which change occurs in the context of stable industry structure; DC required in high velocity environments, in which industry structure is blurring, are simple, experiential unstable processes that rely on quickly created new knowledge and iterative execution to produce adaptive, but unpredictable outcomes (Eisenhardt and Martin 2000).

The DCV argues that firms must be able to reconfigure their resource base in commensuration with the dynamism and changing environmental context to thrive. DC helps organisations develop new sources of competitive advantage or reconfigure existing sources to sustain competitive advantage in an unstable environment.

The current business landscape is characterised by innovation-based competition, waves of creative destruction of existing technological solutions, and economic headwinds (Teece et al. 1997); and is constantly changing. It is therefore difficult to sustain a competitive advantage without the capacity to constantly reconfigure resources and capabilities to fit and align with the changing environment (Helfat et al. 2007). DC are thus essential to the survival and thriving of firms existing in turbulent and unstable environments (Vanpoucke et al. 2014). The value of DCs lie in the created outcomes of resource configurations, and not in the DCs themselves; that is though DC are required, on their own they are not enough conditions for achieving competitive advantage (Eisenhardt and Martin 2000).

## 3 RESEARCH MODEL DEVELOPMENT

Here, we apply RBT and the DCV to the ISM discipline which is responsible for the management of all cybersecurity activities within the organisation (Maynard et al. 2018).

## 3.1 Dynamic Information Security Management Capability

Capability is an organisation's capacity to utilise competences to accomplish organisational goals through focused investment (Peppard and Ward 2004). Capabilities comprise two key aspects which are competences and practices (Bekmamedova and Shanks 2014). Therefore, at the enterprise level, we argue that the ISM capability is expressed as the way ISM competences and ISM practices are strategically utilised to achieve organisational goals. It is worth highlighting that all organisations have some form of ISM capability. For some, however, it may be weak and severely affects the organisation's ability to induce trust and assurance in their business processes by ensuring the protection of their information, and their ability to respond accordingly to information security related threats and strategic change. Those with a strong ISM capability can both build secure and trust inducing business processes for business advantage and respond rapidly to threats and changes in the environmental context of the organisation.

While the RBT is applicable in a stable environment, DC are required when the environment becomes turbulent, unstable and unpredictable. Nowadays, not only must organisations contend with a dynamic business landscape characterised by competition and economical challenges; they must also contend with an equally dynamic cybersecurity landscape that is fraught with highly sophisticated threats. The Dynamic ISM Capability (DISMC) is therefore the organisation's capacity to 1) strategically utilise ISM Competences and ISM Practices to achieve organisational goals and 2) to reconfigure existing ISM resources, competences, practices and capabilities as required in the event of turbulence.

### 3.1.1 ISM Competences

Competences represent an organisation's ability to develop, mobilise and utilise its resources, usually in combination to achieve a given task (Peppard and Ward 2004). In the field of Information Security, examples of resources may include physical and technical assets such firewalls, endpoint protection systems; however, examples of resources specific to the field of ISM will typically be ISM knowledge, experience and skills resident in (or available to) the organisation. Ability to ensure a firewall is effectively configured is likely more valuable than the actual piece of hardware itself.

ISM competences represent the organisation's ability to utilise ISM resources to achieve given task. These competences operate at a departmental or functional level rather than individual. Competences are a cumulation of the skills, knowledge and experience. ISM Skills represent the know-how of ISM – the ability to do; ISM Knowledge represents the know-what – the ability to understand ISM; and ISM Experience represents actual practical involvement in the ISM discipline.

The Dynamic ISM Capability manifests itself at the enterprise level, and it is the organisation's capacity to utilise its ISM competences to accomplish desired goals. The more developed and matured the ISM competences, the more the capacity to achieve organisational goals.

> Proposition 1: there is a positive relationship between ISM Competences and Dynamic ISM Capability

### 3.1.2 ISM Practices

Practices, in contrast to competences, represent how organisation do what they do. Emphasis is on the how and not the ability to do the what. The DCV differentiates between operational routines and strategic routines. While the operational routines represent the day-to-day operations or practices of an organisation. An organisation will have established routines for its various business functions, i.e. routines for recruitment of new staff, routines for providing new computer systems to resuming employees, routines for dealing with suppliers, routines for applying Microsoft's monthly patches and security updates, etc. these routines are known as practices and are usually codified in organisational processes.

Strategic routines, on the other hand, are the established routines for reconfiguring existing operational routines and/or resources to respond to changes in the prevailing environmental context. These strategic routines are known as DC. In a relatively stable environment, the DISMC may only comprise of 'operational routines', however, in a turbulent environment, DC come into effect.

> Proposition 2: there is a positive relationship between ISM Practices and Dynamic ISM Capability

ISM Practices are influenced by the attitude and behaviour of the people that make up the departmental or functional units. ISM Behaviour represents the expression of feelings, action or inaction regarding ISM and ISM Attitude represents the predisposition of the individuals that make up those functional units to certain ideas, values, people, systems, etc., regarding ISM.

## 3.2 Perceived Organisational Value Derived from ISM

Determining the value of ISM within organisations has always been problematic. The public or an interested party's perception of a firm's ISM posture may affect their decision to do business with the firm and this may in turn translate to sales and profit for the organisation. However, it is hard to attribute a figure of these sales to ISM.

Researchers have used different constructs to represent firm performance, such as competitive advantage and sustained competitive advantage. Furthermore, researchers have also used a range of indices to measure firm performance in empirical studies, most of which have been hard financial measurements such as profit, return on investment, sales, etc. (Singh et al. 2016; Wall et al. 2004). Some studies, however, have used other metrics such as behavioural indices and executives' perception of value as viable measurements of firm performance (Tallon et al. 2000). Executives' perception of value has been shown to correlate with actual measured value (Tallon 2014; Tallon and Kraemer 2007). We therefore argue that the executives' perception of the value of ISM would therefore correlate with actual value. Therefore, by gauging and measuring their perception of the value of ISM, we can thus attempt to empirically link ISM capability to firm performance.

The degree to which DISMC contributes to firm performance relies on the strategy and investment decisions made by the organisation (Peppard and Ward 2004). Although cyber security poses a significant concern to all organisations, the extent to which organisations factor cybersecurity into their overall strategy and fund cybersecurity initiatives will determine how DISMC will ultimately affect firm performance. In this study, we have adopted executives' perception of organisational value derived from ISM as a construct to represent firm performance.

> Proposition 3: Dynamic ISM Capability is positively related to Perceived Organisational Value derived from ISM

### 3.3 Effect of Turbulence

In ISM, turbulence refers to the transient nature of the cybersecurity threat landscape such that security controls put in place yesterday may no longer be effective today due to threat evolution. A prevailing turbulent cybersecurity threat environment necessitates a shift of emphasis from preventative controls to a mindset of strategic response (Maynard et al. 2018); it requires developing the organisation's DISMC. The more developed the DISMC, the more the organisation can strategically respond to turbulence, and subsequently, the more the impact on firm performance.

> Proposition 4: Turbulence moderates the influence of Dynamic ISM Capability on Perceived Organisational Value of ISM

### 3.4 Research Model and Summary of Propositions

In figure 1, we present the complete Dynamic ISM Capability model, showing the relationships and the corresponding propositions.

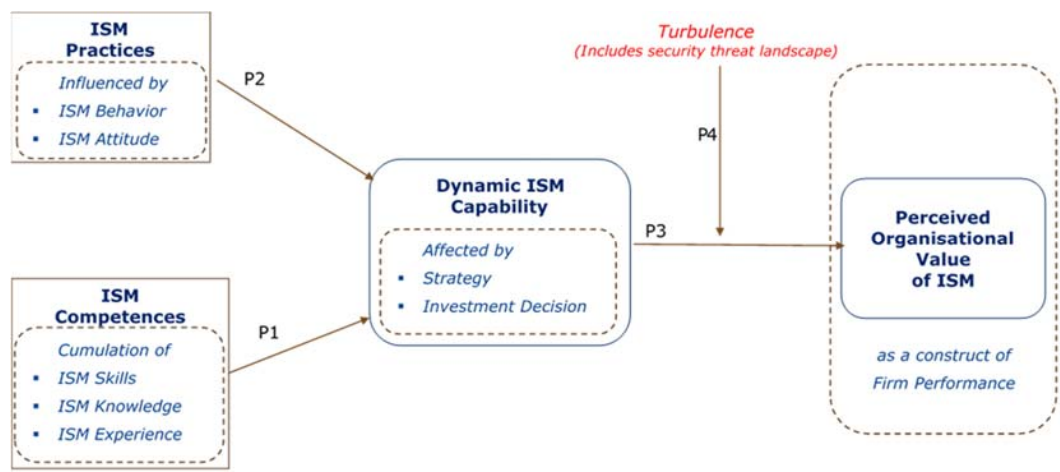

*Figure 1: Dynamic ISM Capability Model*

Proposition 1: there is a positive relationship between ISM Competences and Dynamic ISM Capability

Proposition 2: there is a positive relationship between ISM Practices and Dynamic ISM Capability

Proposition 3: Dynamic ISM Capability is positively related to Perceived Organisational Value derived from ISM

Proposition 4: Turbulence moderates the influence of Dynamic ISM Capability on Perceived Organisational Value of ISM

## 4 METHOD

We will develop a detailed survey questionnaire, based upon the proposed model. As there are no previous questionnaire-based contributions in the ISM literature, it is not possible to adapt specific questions and item measures from existing literature. Consequently, once we have developed a draft questionnaire, it will be necessary to subject it to a rigorous validation process.

The draft survey questionnaire will be validated through a phased pre-testing regime: firstly, with a small group of experienced IS researchers and academics; then, after refinement, we will test it with a small group of business, IT and Information Security leaders. After completing refinement of the questionnaire, we will administer the full survey to selected business, IT and information security leaders from medium to large organisations. The table 1 below shows the representative job titles of expected participants in this study.

| Business Leadership | IT leadership | ISM Leadership |
| --- | --- | --- |
| CEO | CIO | CISO/CSO |
| Directors | IT Directors | Information Security Manager |
| GMs | IT Manager | |

*Table 1. Representative Job Titles of Proposed Survey Respondents*

## 5  CONCLUSION AND EXPECTED CONTRIBUTIONS

The organisational ISM function must adopt a strategic perspective and be able to function in the face of a dynamic threat landscape to help the organisations to achieve their goals. This research sets out to 1) define and describe an organisational level dynamic ISM capability, 2) to develop a strategic model that links resources with this dynamic capability, and then 3) empirically demonstrate how it contributes to firm performance. We found little or no articulation of dynamic ISM capability with its link to organisational performance. We therefore drew concepts from RBT and DCV to define, describe and develop a conceptual model of the Dynamic ISM Capability. In the future, we will empirically test this model to determine causality between dynamic ISM capability and organisational performance.

This study contributes to theory. Firstly, the proposed variance model strives to empirically demonstrate causality between ISM capability and firm performance, which is rarely discussed in ISM literature. Secondly, it extends existing ISM domain knowledge by drawing concepts from RBT and DCV and interpreting them into the ISM literature.

This study contributes to practice. Firstly, the proposed framework can help organisations recognise and develop their dynamic ISM capability, which eventually leads to better firm performance. Secondly, empirically demonstrated causality between ISM capability and firm performance can be used by organisations to strengthen the justification for investment in their ISM capability.

## 6  References

Ahmad, A., Hadjkiss, J., and Ruighaver, A. B. 2012. "Incident Response Teams - Challenges in Supporting the Organizational Security Function," *Computers & Security* (31:5), pp. 643-652.

Ahmad, A., Maynard, S. B., and Park, S. 2014. "Information Security Strategies: Towards an Organizational Multi-Strategy Perspective," *Journal of Intelligent Manufacturing*).

Ahmad, A., Maynard, S. B., and Shanks, G. 2015. "A Case Analysis of Information Systems and Security Incident Responses," *International Journal of Information Management* (35:6), pp. 717-723.

Ahmad, A., Webb, J., Desouza, K. C., and Boorman, J. 2019. "Strategically-Motivated Advanced Persistent Threat: Definition, Process, Tactics and a Disinformation Model of Counterattack," *Computers & Security*).

Barney, J. 1991. "Firm Resources and Sustained Competitive Advantage." Sage Publications, Inc.

Baskerville, R., Spagnoletti, P., and Kim, J. 2014. "Incident-Centered Information Security: Managing a Strategic Balance between Prevention and Response," *Information & Management* (51:1), pp. 138-151.

Bekmamedova, N., and Shanks, G. 2014. "Social Media Analytics and Business Value: A Theoretical Framework and Case Study." IEEE.

Dehning, B., and Stratopoulos, T. 2003. "Determinants of a Sustainable Competitive Advantage Due to an It-Enabled Strategy," *The Journal of Strategic Information Systems* (12:1), pp. 7-28.

Desouza, K. C., Ahmad, A., Naseer, H., and Sharma, M. 2020. "Weaponizing Information Systems for Political Disruption: The Actor, Lever, Effects, and Response Taxonomy (Alert)," *Computers & Security*), p. 101606.

Eisenhardt, K. M., and Martin, J. A. 2000. "Dynamic Capabilities: What Are They?," *Strategic Management Journal* (21:10/11).

Grant, R. M. 1996. "Toward a Knowledge-Based Theory of the Firm." Great Britain: JOHN WILEY & SONS LTD.

Helfat, C. E., Finkelstein, S., Mitchell, W., Peteraf, M. A., and Singh, H. 2007. *Dynamic Capabilities : Understanding Strategic Change in Organizations*. Malden, Mass. [u.a.].

Huang, K.-F., Dyerson, R., Wu, L.-Y., and Harindranath, G. 2015. "From Temporary Competitive Advantage to Sustainable Competitive Advantage," *British Journal of Management*:4).

Maynard, S. B., Onibere, M., and Ahmad, A. 2018. "Defining the Strategic Role of the Chief Information Security Officer," *Pacific Asia Journal of the Association for Information Systems* (10:3).

Peppard, J., and Ward, J. 2004. "Beyond Strategic Information Systems: Towards an Is Capability," *Journal of Strategic Information Systems* (13), pp. 167-194.

Pisano, G. P. 1994. "Knowledge, Integration, and the Locus of Learning: An Empirical Analysis of Process Development," *Strategic Management Journal*:SPEISS).

Schiavone, S., Garg, L., and Summers, K. 2014. "Ontology of Information Security in Enterprises," *Electronic Journal Information Systems Evaluation Volume* (17:1).

Singh, S., Darwish, T. K., and Potočnik, K. 2016. "Measuring Organizational Performance: A Case for Subjective Measures," *British Journal of Management* (27:1), pp. 214-224.

Smiraus, M., and Jasek, R. 2011. "Risks of Advanced Persistent Threats and Defense against Them," *Annals of DAAAM & Proceedings*).

Sood, A. K., and Enbody, R. J. 2013. "Targeted Cyberattacks: A Superset of Advanced Persistent Threats," *IEEE Security & Privacy Magazine* (11:1).

Tallon, P. P. 2014. "Do You See What I See? The Search for Consensus among Executives' Perceptions of It Business Value." pp. 306-325.

Tallon, P. P., and Kraemer, K. L. 2007. "Fact or Fiction? A Sensemaking Perspective on the Reality Behind Executives' Perceptions of It Business Value," *Journal of Management Information Systems* (24:1), pp. 13-54.

Tallon, P. P., Kraemer, K. L., and Gurbaxani, V. 2000. "Executives' Perceptions of the Business Value of Information Technology: A Process-Oriented Approach," *Journal of Management Information Systems* (16:4), pp. 145-173.

Tankard, C. 2011. "Advanced Persistent Threats and How to Monitor and Deter Them," *Network Security*:8).

Teece, D. J. 2000. "Strategies for Managing Knowledge Assets: The Role of Firm Structure and Industrial Context." Great Britain: PERGAMON.

Teece, D. J., Pisano, G., and Shuen, A. 1997. "Dynamic Capabilities and Strategic Management." Great Britain: JOHN WILEY & SONS LTD.

Vanpoucke, E., Vereecke, A., and Wetzels, M. 2014. "Developing Supplier Integration Capabilities for Sustainable Competitive Advantage: A Dynamic Capabilities Approach," *Journal of Operations Management* (32:7), pp. 446-461.

Wall, T. D., Michie, J., Patterson, M., Wood, S. J., Sheehan, M., Clegg, C. W., and West, M. 2004. "On the Validity of Subjective Measures of Company Performance." pp. 95-118.